\documentclass[letterpaper, 10pt, conference]{ieeeconf}
\usepackage{hyperref}
\usepackage{graphicx}
\usepackage{amsmath}
\usepackage{amssymb}
\usepackage{float}
\usepackage{array}
\usepackage{booktabs}
\usepackage{url}
\usepackage[inkscapelatex=false]{svg} 

\usepackage{graphicx}

\DeclareMathOperator*{\argmin}{arg\,min}
\IEEEoverridecommandlockouts
\allowdisplaybreaks
\usepackage{soul,color}

\newcommand{\cK}{\mathcal{K}}

\newcommand{\bE}{\mathbf{E}}
\newcommand{\bg}{\mathbf{g}}
\newcommand{\bH}{\mathbf{H}}

\newcommand{\bF}{\mathbf{F}}

\title{\textbf{Polynomial Parametric Koopman Operators for Stochastic MPC}}

\author{{Efstathios Iliakis}$^{1*}$, {Wallace Gian Yion Tan}$^{1*}$, Liang Wu$^{2}$, J\'an Drgo\v na$^{2}$, and {Richard D. Braatz}$^{1\dag}$
\thanks{$^*$Authors contributed equally. $^1${Efstathios Iliakis}, {Wallace Gian Yion Tan}, and {Richard D. Braatz} are with the 
Massachusetts Institute of Technology. $^2${Liang Wu and  J\'an Drgo\v na} are with 
Johns Hopkins University.
$^{\dag}$Corresponding author: Richard D. Braatz, email: {\tt\small braatz@mit.edu}.}
\thanks{This research was supported by the U.S. Food and Drug Administration under the FDA BAA-22-00123 program, Award Number 75F40122C00200, and by the U.S. DOE, Office of Science, ASCR program under the Scientific Discovery through Advanced Computing (SciDAC) Institute “LEADS: LEarning-Accelerated Domain Science.”  
{Efstathios Iliakis was further supported by the Onassis Foundation Scholarships' Program.} {Wallace Gian Yion Tan} was further supported by the MathWorks Chemical Engineering Fellowship.} 
\thanks{ChatGPT was used for minor grammatical corrections.}
}

\begin{document}
	
	\maketitle 
	
\begin{abstract}
This paper develops a parametric Koopman operator framework for Stochastic Model Predictive Control (SMPC), where the Koopman operator is parametrized by Polynomial Chaos Expansions (PCEs). The model is learned from data using the Extended Dynamic Mode Decomposition -- Dictionary Learning (EDMD-DL) method, which preserves the convex least-squares structure for the PCE coefficients of the EDMD matrix. Unlike conventional stochastic Galerkin projection approaches, we derive a condensed deterministic reformulation of the SMPC problem whose dimension scales only with the control horizon and input dimension, and is independent of both the lifted state dimension and the number of retained PCE terms. Our framework, therefore, enables efficient nonlinear SMPC problems with expectation and second-order moment constraints with standard convex optimization solvers. Numerical examples demonstrate the efficacy of our framework for uncertainty-aware SMPC of nonlinear systems. 
\end{abstract}




 \vspace{-3pt}

 \section{Introduction}
Many real-world control systems operate under significant parametric uncertainties arising from uncertain physical parameters and varying operating conditions. Some common examples include chemical reaction systems with uncertain kinetic and heat transfer coefficients \cite{ruppen1995optimization} and mechanical systems with uncertain mass or stiffness parameters \cite{mammarella2018offline}. In such scenarios, stochastic model predictive control (SMPC) has emerged as a principled approach to handle uncertainty by explicitly incorporating probabilistic models of uncertain parameters into the predictive control framework \cite{mesbah2016stochastic}. 

A fundamental challenge in the practical implementation of SMPC is the need to propagate uncertainty through the system dynamics and evaluate expectations appearing in the cost and constraint functions. When the underlying system dynamics are nonlinear, the resulting SMPC problem is typically formulated as a nonlinear program (NLP), which must be solved at each sampling time. Such NLPs are generally computationally expensive and difficult to integrate with real-time optimization in practical applications. In contrast, for systems with linear dynamics and parametric uncertainty, a rich set of computationally efficient methods exists in the literature, including approaches based on polynomial chaos expansions (PCEs) \cite{tan2025offset,dai2016distributed,bernardini2011stabilizing,mishra2024polynomial}. These methods approximate the stochastic dynamics by projecting onto a set of orthogonal polynomials of the parametric uncertainty, which results in a deterministic expanded system of PCE coefficients that can be controlled with deterministic MPC methods. 

Recently, a popular data-driven approach to improve the computational tractability of nonlinear deterministic MPC is to learn a linear high-dimensional representation of the dynamics with Koopman operator methods~\cite{korda2018linear,proctor2018generalizing}. These methods enable the reformulation of nonlinear MPC optimization problems as compact QPs after condensation to remove the high-dimensional state dependence. To address modeling errors arising from the finite-dimensional approximation of the Koopman operator, robust tube-based MPC formulations have been proposed that ensure closed-loop stability under bounded additive disturbances~\cite{zhang2022robust}, while stochastic MPC has been combined with Koopman models to handle approximation uncertainty probabilistically through chance constraints~\cite{kim2025ksmpc}. However, neither approach incorporates parametric uncertainty directly into the Koopman representation itself. Parametric Koopman operators have been proposed to extend Koopman representations to systems with parametric inputs~\cite{guo2025learning,williams2016extending}, which implicitly suggests parameterizing the Koopman operator by a PCE. This approach enables the application of linear SMPC methods to systems with nonlinear dynamics and parametric uncertainty, which is the gap addressed by our proposed Polynomial Parametric Koopman Operators (PPKO) framework.

\subsection{Contributions}
The main contributions of this article are
\begin{itemize}
\item We propose a PCE-based parametric Koopman operator framework for nonlinear dynamical systems with time-invariant parametric uncertainty. 
\item We derive a condensed deterministic reformulation of the SMPC problem whose dimension scales only with the control horizon and input dimension, independent of the number of Koopman observables and PCE terms.
\item Numerical examples on the Duffing oscillator and the control of product concentration in a chemical reaction network demonstrate the effectiveness of our method on nonlinear systems with strong parametric uncertainty. 
\end{itemize}


\section{Problem Formulation}
Consider the problem of regulating nonlinear stochastic discrete-time systems to the origin:
\begin{equation}\label{eq:nonlinear}
 x_{t+1} = f(x_t,u_t,\theta),
\end{equation}
where ${x}_t \in \mathbb{R}^{n_x}$ are the states and $u_t \in \mathbb{R}^{n_u}$ are the control actions at time $t$, respectively, and $\theta$ is a random vector denoting the parametric uncertainty. Stochastic Model Predictive Control (SMPC) addresses this problem by solving a finite-horizon stochastic optimal control problem with additional state and control constraints. Assuming that a measurement of the state $x_0$ is available at time $t = 0$, the finite-horizon chance-constrained SMPC problem with a linear quadratic regulator (LQR) type cost is 
\noindent\rule{\columnwidth}{0.4pt}
\textbf{Stochastic MPC}
{\small
\begin{subequations}\label{eq:smpc_problem}
\begin{align}
\min_{\{u_t\}_{t=0}^{H-1}} 
&~ \sum_{t=0}^{H-1} 
\mathbb{E}_\theta\big[x_t^\top Q x_t + u_t^\top R u_t \big]
+ \mathbb{E}_\theta\big[x_H^\top Q_f x_H\big] \label{eq:smpc-cost}
\\
\text{s.t.}~
& x_{t+1} = f(x_t,u_t,\theta),\quad t=0,\dots,H-1, \label{eq:smpc-dyn}
\\
& x_t^{\min} \le \mathbb{E}[x_t] \le x_t^{\max},
\quad t=1,\dots,H,
\\
& \mathbb{E}\big[(a_t^\top x_t-b_t)^2\big] \!\le c_{t}^2,\quad
 t=1,\dots,H,\label{eq:2ndorderconstraints}
\\
& u_{\min} \le u_t \le u_{\max},\quad t=0,\dots,H -1 
\end{align}
\end{subequations}
}
\noindent\rule{\columnwidth}{0.2pt}
where $H$ denotes the prediction horizon; the matrices $Q \succeq 0$, $R \succ 0$, and $Q_f \succeq 0$ are weighting matrices; and the vectors $u_{\min}, u_{\max}, \in \mathbb{R}^{n_u}$ define the input bounds. The vectors \(a_t \in \mathbb{R}^{n_x}\), \(t=1,\ldots,m\), define the directions
associated with the second-order safety/risk constraints, while \(t=1,\ldots,H\)
selects the time instant along the prediction horizon at which each constraint is enforced.
The scalar \(b_t\) specifies the corresponding offset, and \(c_{t}\) specifies the allowable
second-moment bound for the constraint at time \(t\). The second-moment constraints \eqref{eq:2ndorderconstraints} can be viewed as being convex relaxations of probabilistic constraints via the Chebyshev inequality \cite{paulson2020stochastic}.

The nonlinear dynamics \eqref{eq:smpc-dyn} in SMPC \eqref{eq:smpc_problem} usually results in an optimization that is both nonlinear and nonconvex.
Consequently, solving SMPC \eqref{eq:smpc_problem} online becomes computationally expensive, posing a major hurdle for fast real-time applications.

In deterministic nonlinear MPC, Koopman operators have been integrated with MPC formulations to lift the nonlinear dynamics in \eqref{eq:smpc-dyn} to a higher dimensional space of observables, where the lifted states evolve linearly in time. 
This lifting enables the reformulation of the nonlinear MPC problem as a linear MPC problem. 
With quadratic stage costs and only linear constraints, the resulting optimization problem is a convex quadratic program (QP). In addition to using standard QP solvers, solving a QP exhibits the same $O(n^3)$ time complexity as solving a linear system of equations  \cite{11431115}, and further offers data-independent execution time certificates \cite{10683964,11240592}, which is required for real-time control applications. When second-order moment constraints are included, the lifted formulation instead leads to a convex quadratically constrained quadratic program (QCQP), which can be solved efficiently using modern conic optimization solvers (e.g., CLARABEL). 
\section{Polynomial Chaos Theory}\label{section:PCE}
This section briefly reviews polynomial chaos theory for propagating parametric uncertainty in dynamical systems. Let $\theta$ be a real-valued random variable with a sufficiently regular probability density function $f_\theta(x)$ (e.g., normal, uniform, or beta). Any square-integrable function $p(\theta)$ admits a PCE representation in terms of polynomial basis functions and corresponding coefficients. In particular, any square-integrable $p(\theta)$ can be expanded with respect to $\theta$ as
\begin{equation}\label{eq:basic-PCE}
p(\theta) = \sum_{k=0}^\infty p_k \phi_k (\theta),
\end{equation}
where $\phi_k$ is the $k$th polynomial basis function associated with $\theta$, and $p_k$ is the corresponding PCE coefficient. The series in \eqref{eq:basic-PCE} converges in the $\mathcal{L}_2$ sense, and the polynomials $\{\phi_k\}$ form a complete orthonormal basis, i.e., $\langle \phi_i(\theta), \phi_j(\theta) \rangle = \delta_{i,j}$. The inner product is defined by the density-weighted integral $\langle F(\theta), G(\theta) \rangle := \int_{\mathbb{R}^N} F(x)G(x) f_\theta (x)\,dx$, where $f_\theta (x)$ is the weight function depending on the distribution of $\theta$. The coefficients are obtained by projection: $p_k = \langle p(\theta), \phi_k(\theta) \rangle$.

Some canonical polynomial basis functions for various probability distributions are given in Table \ref{table: PCE}. Due to the orthonormality of $\phi_k$,
\begin{equation}\label{eq: PCE moments}
\mathbb{E}[p(\theta)] = p_0, \qquad \text{Var}[p(\theta)] = \sum_{k=1}^\infty (p_k)^2.
\end{equation}
In a higher dimensional setting where $\mathbf{\theta} \in \mathbb{R}^{d}$ consists of independent random variables $\theta_i$ for $d \geq 2$, the Polynomial Chaos Expansion (PCE) basis functions are constructed as the tensor products of univariate orthonormal polynomials. For instance, if $\mathbf{\theta}$ is uniformly distributed on the hypercube $[-1,1]^d$, the multivariate basis functions $\Phi_{\alpha}(\mathbf{\theta})$ are defined as the product of Legendre polynomials $\phi_{\alpha_i}(\theta_i)$ in each coordinate. The general expansion is given by
\begin{equation}
    p(\mathbf{\theta}) = \sum_{\alpha \in \mathbb{N}^d} p_\alpha \Phi_{\alpha}(\mathbf{\theta}),
\end{equation}
where $\Phi_{\alpha}(\mathbf{\theta}) = \prod_{i=1}^d \phi_{\alpha_i}(\theta_i)$, and $\alpha = (\alpha_1, \dots, \alpha_d)$ is a multi-index representing the polynomial degree in each dimension. In practice, this infinite series must be truncated to a finite set of terms. We employ the total degree (TD) approximation, which restricts the expansion to multi-indices whose $L_1$-norm does not exceed a maximum degree $D$,
\begin{equation}
    p(\mathbf{\theta}) \approx \sum_{|\alpha|_1 \le D} p_\alpha \Phi_{\alpha}(\mathbf{\theta}),
\end{equation}
where $|\alpha|_1 = \sum_{i=1}^d \alpha_i$ denotes the total degree. For computational simplicity, after re-ordering the multivariate basis functions $\{\Phi_{\alpha}(\mathbf{\theta})\}_{|\alpha|_1 \le D}$ (e.g., using lexicographic order), we denote them as $\{\Phi_k(\mathbf{\theta})\}_{k=0}^{N_D-1}$, where $N_D = \binom{d+D}{D}$ represents the number of polynomials in the truncated expansion.
\begin{table}[h!]
\centering

    \vspace{0.2cm}
    
\caption{Some canonical choices of orthogonal polynomial bases for PCEs. Theoretical details may be found in \cite{xiu2002wiener}.}
\label{table: PCE}
\vspace{-0.2cm}
\begin{tabular}{@{}ccc@{}} 
 \hline Distribution $f_\theta (x)$ & Support $x$ & Polynomial Basis Functions \\ [0.5ex] 
 \hline\hline
Uniform & $[-1,1]$ & Legendre $P_n(x)$ \\
Exponential & $\mathbb{R}_{\geq 0}$  & Laguerre $L_n(x)$  \\
Gaussian & $\mathbb{R}$ & Hermite $H_n(x)$ \\
 \hline
\end{tabular}
\vspace{-0.4cm}
\end{table}

\section{Koopman Operators for Nonlinear Dynamical Systems}
This section briefly reviews conventional Koopman operators for nonlinear dynamical systems and summarizes the EDMD-DL approach for learning finite-dimensional approximations from data \cite{li2017extended}. Consider an autonomous discrete-time nonlinear system $x_{t+1} = f(x_t)$, where $f : X \to  X$, where $X \in \mathbb{R}^{n_x}$ denotes the state space. The Koopman operator $\cK$ provides a linear representation of the dynamics by acting on the space of observable functions $\psi : X \to \mathbb{R}$, $\mathcal{K} \psi := \psi\circ f$. The Koopman operator acts on an infinite-dimensional separable Hilbert space and must be truncated in practice, and many finite-dimensional approximations have been proposed in the literature. An increasingly popular approach is Extended Dynamic Mode Decomposition with Dictionary Learning (EDMD-DL), which utilizes a trainable architecture (usually a neural network) to learn an optimal finite dictionary of observables. This approach allows for the construction of a least-squares linear model in a lifted space that specifically minimizes the reconstruction and prediction errors of the nonlinear system.

\subsection{Extended Dynamic Mode Decomposition with Dictionary Learning}
EDMD is a data-driven algorithm that approximates the Koopman operator by restricting its action to the span of a finite dictionary of observables $\{\psi_i\}_{i=1}^{n_\psi}$. Denoting $\Psi(x) := \left[\psi_1(x),\dots,\psi_{n_\psi}(x)\right]^\top$ and the lifted state $z_k := \Psi(x_k)$, EDMD identifies a matrix $K \in \mathbb{R}^{n_\psi \times n_\psi}$ such that $z_{k,+} \approx K z_k$.

Specifically, given snapshot pairs $\{(x_j,x_{j,+})\}_{j=1}^{M}$ generated by the update mapping $x_{j,+}=f(x_j)$, the approximated matrix  $K$ (termed as the Koopman matrix) is identified by least squares:
\begin{equation}\label{eq:edmd_ls}
K = \argmin_{\tilde K\in\mathbb{R}^{n_\psi\times n_\psi}} \frac1M \sum_{j=1}^{M} \left\| \Psi(x_{j,+}) - \tilde K\,\Psi(x_j) \right\|_2^2.
\end{equation}
In EDMD-DL, the dictionary $\Psi(x) = \Psi(x; w_\psi)$ is parametrized by neural network weights $w_\psi$, which can be chosen as a conventional feedforward neural network. This approach helps in finding a suitable dictionary that is (approximately) invariant under the action of the Koopman operator. EDMD-DL jointly identifies the Koopman matrix $K$ and neural network weights $w_\psi$ by solving the nonlinear program
\begin{equation}\label{eq:edmd_dl}
\min_{K\in \mathbb{R}^{n_\psi\times n_\psi},\; w_\psi}\; \frac{1}{M}\sum_{j=1}^{M} \left\| \Psi(x_{j,+}; w_\psi) - K\,\Psi(x_j; w_\psi) \right\|_2^2.
\end{equation}
The learned dictionary $\Psi(x; w_\psi)$ is typically constrained to include basic observables such as the coordinate functions $\psi_i(x) := 1$ and $\psi_i(x) := x_i$ for $i \in [1,n_x]$, which helps avoid the trivial solution $\Psi(x) \equiv 0$ in \eqref{eq:edmd_dl}. In addition, regularization (e.g., $\ell_1$ or $\ell_2$ penalties) is often incorporated to promote sparsity in $K$ and/or improve numerical stability. Problem \eqref{eq:edmd_dl} can be solved using standard machine-learning optimizers (e.g., Adam) or via alternating minimization, which iterates between optimizing $K$ with $w_\psi$ fixed and optimizing $w_\psi$ with $K$ fixed. 
\section{Polynomial Parametric Koopman Models}
This section describes an extension of conventional Koopman Models for parametric uncertainty, which we term as {\em Polynomial Parametric Koopman Operators} (PPKOs). As stated in the introduction, due to the presence of parametrized nonlinearity in \eqref{eq:smpc-dyn}, standard Koopman operator formulations may fail to provide efficient and distribution-aware predictions necessary for solving \eqref{eq:smpc_problem}, which motivates the formulation of parametric Koopman models. Parametric Koopman Models were first formulated in \cite{guo2025learning} for constructing Koopman approximations for dynamical systems with unknown parameters. 

We begin by reviewing the parametric Koopman formulation in \cite{guo2025learning} for autonomous systems with some minor modifications; extensions to controlled systems are presented later in this article. Let $\theta \in \mathbb{P} \subseteq\mathbb{R}^{p}$ denote a vector of (unknown and/or uncertain) parameters, where $p$ is the number of the parameters.
Consider parametrized autonomous discrete-time dynamics of the form
\begin{equation}\label{eq:param_dyn}
 x_{t+1} = f(x_t,\theta),
 \end{equation}
where $f$ is a (jointly) continuous function.
As in the standard Koopman setting, introduce a space of observable functions $\psi : X \to \mathbb{R}$.
Given an observable $\psi$, the dynamics \eqref{eq:param_dyn} induces the time evolution $\psi(x_{k+1}) = \psi(f(x_k,\theta))$, which depends on the parameter $\theta$, which allows the definition of a parametrized Koopman operator $\cK(\theta)$ by
\begin{equation}\label{eq:param_koopman_def}
\cK(\theta)\psi(x) := \psi \circ f (x, \theta).
\end{equation}
For a fixed parameter realization $\theta$, the Koopman operator acts linearly on observables; that is, for any observables $\psi_1,\psi_2$ and $a,b \in \mathbb{R}$,
\begin{equation}\label{eq:param_koopman_lin}
\cK(\theta)\big(a\psi_1 + b\psi_2\big) = a\,\cK(\theta)\psi_1 + b\,\cK(\theta)\psi_2.
\end{equation}

 \subsection{Polynomial Parameter Dependence of the Koopman Matrix}
Analogous to the deterministic case with no parametric uncertainty, for each parameter realization $\theta$, the Koopman operator $\cK(\theta)$ is infinite-dimensional and must be truncated for numerical approximation. The approach for learning a parametric Koopman operator in \cite{guo2025learning} is as follows. Similar to the previous section, denoting $\Psi(x) := \left[\psi_1(x),\dots,\psi_{n_\psi}(x)\right]^\top$ and the lifted state $z_t := \Psi(x_t)$, the aim is to identify a matrix-valued function $K(\theta) \in \mathbb{R}^{n_\psi \times n_\psi}$ such that $z_{t,+} \approx K(\theta) z_t$ for all $\theta$. An important distinction between their approach and our approach is the parametrization method. In their approach, the Koopman matrix is parametrized by a deep neural network, which provides sufficient expressivity to capture the dynamics over the range of possible parameters. In our approach, we enforce that the Koopman matrix is parametrized by a PCE, which enables fast state uncertainty quantification. Specifically, assume that $K(\theta)$ is defined by a truncated PCE,
\begin{equation}\label{eq:PPKO}
K(\theta) = \sum_{k =0}^{N_K-1} K_k \phi_k (\theta)
\end{equation}
where $K_k \in \mathbb{R}^{n_\psi \times n_\psi}$ are the PCE coefficient matrices and $\phi_k (\theta)$ is the kth component of the (multivariate) PCE polynomial-basis. We refer to the representation in \eqref{eq:PPKO} as a {\em polynomial parametric Koopman operator} (PPKO). Analogous to conventional EDMD, we then learn the matrices $\{K_k\}$ and dictionary weights $w_\psi$ from snapshot pairs $\{(x_j,x_{j,+},\theta_j)\}_{j=1}^{M}$ by minimizing the least-squares loss,
\begin{small}
      \begin{equation}\label{eq:ppko_ls}
\min_{\substack{\{K_k\},\\ w_\psi}}\; \frac{1}{M}\sum_{j=1}^{M}\left\| \Psi(x_{j,+};w_\psi) - \left(\sum_{k =0}^{N_K-1} K_k \phi_k (\theta)\!\right)\!\Psi(x_j;w_\psi)\right\|_2^2.
\end{equation}
\end{small}
\noindent
For a fixed $w_\psi$, the least-squares problem \eqref{eq:ppko_ls} is linear and convex in $\{K_k\}_0^{N_K-1}$ and can be solved at the end of every epoch where, after optimizing $w_\psi$ with fixed $\{K_k\}_0^{N_K-1}$, least square regression is performed to determine the coefficients of matrices $\{K_k\}_0^{N_k-1}$ based on the whole training set. Once the PPKO is identified, for each fixed realization $\theta$, state predictions can be computed via the linear time-invariant (LTI) dynamics in the lifted space,
\begin{equation}\label{eq:LTI}
z_{t+1} = K(\theta)z_{t},~x_{t+1} = Cz_{t+1} ,
\end{equation}
where $C = [0_{1 \times n_\psi}, I_{n_x}, 0_{n_\psi -1-n_x \times n_\psi}]$ is the output matrix (since our observables already contain the states by definition).

\section{PPKO for Controlled Systems}
We now extend the PPKO framework to controlled systems. Consider nonlinear discrete-time systems of the form \eqref{eq:nonlinear} with time-invariant uncertain parameters $\theta$. In this article, for each uncertain parameter realization $\theta$, we approximate the controlled Koopman operator by a control-affine representation in the lifted coordinates $z_t = \Psi(x_t;w_\psi)$: 
\begin{subequations}\label{eq:LTIc}
\begin{align}
z_{t+1}& = A(\theta)z_{t} + B(\theta) u_{t}  \\
x_{t+1}& = Cz_{t+1} 
\end{align}
\end{subequations}
where both $A(\theta)$ and $B(\theta)$ are defined by truncated PCEs:
\begin{equation}
A(\theta) = \sum_{k=0}^{N_K-1} A_k \phi_k (\theta), \quad\quad B(\theta) = \sum_{k=0}^{N_K-1} B_k \phi_k (\theta),
\end{equation}
and can be obtained, analogously to the autonomous EDMD-DL approach, by obtaining snapshot pairs $\{(x_j,u_j,x_{j,+},\theta_j)\}_{j=1}^{M}$ and minimizing the  least-squares loss,
\begin{small}
\begin{equation}\label{eq:ppko_control_ls}
\begin{aligned}
\min_{\substack{\{A_k\},\{B_k\},\\ w_\psi}}\ & \frac{1}{M}\!\sum_{j=1}^{M}\Bigg\| \Psi(x_{j,+};w_\psi)
 - \!\Bigg(\sum_{k=0}^{N_K-1} A_k \phi_k (\theta)\!\Bigg)\Psi(x_j;w_\psi) \\
& \hspace{5.3em} - \!\Bigg( \sum_{k=0}^{N_K-1} B_k \phi_k (\theta)\!\Bigg)u_j \Bigg\|_2^2,
\end{aligned}
\end{equation}
\end{small}
where $u_j$ denotes the input applied in snapshot $j$ and $x_{j,+}$ is the successor state. The identified PPKO \eqref{eq:LTIc} enables uncertainty propagation for the nonlinear model \eqref{eq:nonlinear} in lifted coordinates, and is used for deterministic reformulation of the SMPC problem \eqref{eq:smpc_problem} in the next section. 

\section{Condensed Deterministic SMPC Reformulation}
%
%
%
%
%
%
%
Once a PCE LTI model such as \eqref{eq:LTIc} is identified, the conventional approach for deterministic SMPC reformulation is to apply Stochastic Galerkin Projections (SGPs) to \eqref{eq:LTIc}, which results in a deterministic expanded LTI system for the PCE coefficients. The conventional SGP approach increases the state dimension proportionately with the number of retained PCE polynomials, which increases the computational burden of the deterministic reformulation of SMPC. SGPs also induce additional truncation error during uncertain propagation and the errors may propagate along the prediction horizon. In this section, inspired by condensing methods in MPC, we propose a deterministic reformulation approach that explicitly avoids the use of SGPs, which eliminates truncation errors and ensures the resulting optimization depends only on the control horizon and input dimension, and does not scale with the dimension of the lifted (PCE or Koopman) state.

First define the stacked vectors $X = [x_1^\top, \cdots{}, x_H^\top]^\top$ and $U = [u_0^\top, \cdots{}, u_{H-1}^\top]^\top$. Recursively applying \eqref{eq:LTIc}, and noting that $z_0 = \Psi(x_0)$, results in 
\begin{equation}
X = \bE(\theta) z_0 + \bF(\theta) U  
\end{equation}
where 
{\scriptsize
\[
\mathbf{E}(\theta) =
\!\begin{bmatrix}
C A(\theta) \\
C A(\theta)^2 \\
\vdots \\
C A(\theta)^H
\end{bmatrix}\!,
\]
\begin{equation}
\label{eq:E_F_def}
\mathbf{F}(\theta) =\!
\begin{bmatrix}
C B(\theta) & 0 & \cdots & 0 \\
C A(\theta) B & C B(\theta) & \cdots & 0 \\
\vdots & \vdots & \ddots & \vdots \\
C A(\theta)^{H-1}B(\theta) & C A(\theta)^{H-2} B(\theta) & \cdots & C B(\theta)
\end{bmatrix}\!\!.
\end{equation}
}
Since $A(\theta)$ and $B(\theta)$ are both polynomial functions of $\theta$, $\bE(\theta)$ and $\bF(\theta)$ both inherit polynomial (and therefore analytic) dependence on $\theta$. Now the SMPC LQR cost \eqref{eq:smpc-cost} can be rewritten as 
\begin{equation}\label{eq:pcecost}
\begin{aligned}
J(U)
&=
\mathbb{E}_{\theta}
\Big[
U^\top \bar{R} U
\\
&\quad +
(\mathbf{E}(\theta)z_0+\mathbf{F}(\theta)U)^\top
\bar{Q}
(\mathbf{E}(\theta)z_0+\mathbf{F}(\theta)U)
\Big]
\end{aligned}
\end{equation}
where $\bar{R}=\mathrm{blkdiag}(R,\cdots{},R)$ and $\bar{Q}=\mathrm{blkdiag}(Q,\cdots{},Q,Q_f)$. This expression can be further condensed into 
\begin{equation}
J(U) = U^\top\bH U
+ 2\, \mathbf{g}^\top U
+ z_0^\top
\mathbb{E}_{\theta}\!\left[
\mathbf{E}^\top \bar Q \mathbf{E}
\right]
\!z_0 ,
\end{equation}
where
$\bH := \bar R
+ \mathbb{E}_{\theta}\!\left[
\mathbf{F}^\top \bar Q \mathbf{F}
\right]$ and 
$
\mathbf{g} := 
\mathbb{E}_{\theta}\!\left[
\mathbf{F}^\top \bar Q \mathbf{E}
\right] \!z_0$.
$\bH$ and $\bg$ can be precomputed accurately using Gaussian quadrature, because of the polynomial dependence of $\bE$ and $\bF$ on $\theta$, so that all uncertainty integration in computing the cost $J(U)$ is performed offline. Furthermore, for the expectation in the linear constraints, defining $\bE_t(\theta)$ and $\bF_t(\theta)$ as the $t$th row of $\bE(\theta)$ and $\bF(\theta)$ respectively, we have that
\begin{equation}
\mathbb{E}_{\theta}[x_t]
=
\mathbb{E}_{\theta}\!\left[
\mathbf{E}_t(\theta) z_0
+
\mathbf{F}_t(\theta) U
\right]
=
\bar{\mathbf{E}}_t z_0
+
\bar{\mathbf{F}}_t U,
\end{equation}
where $\bar{\mathbf{E}}_t
:=
\mathbb{E}_{\theta}
\!\left[
\mathbf{E}_t(\theta)
\right],
~
\bar{\mathbf{F}}_t
:=
\mathbb{E}_{\theta}
\!\left[
\mathbf{F}_t(\theta)
\right].$
For the second moment constraints, defining
\begin{equation}
\eta := \begin{bmatrix}
    z_0\\
    U
\end{bmatrix},
\qquad
\mathbf{G}_t(\theta) := \left[\, \mathbf{E}_t(\theta)\;\; \mathbf{F}_t(\theta)\,\right],
\end{equation}
gives 
\begin{equation}
\mathbb{E}_{\theta}
\!\left[
(a_t^\top x_t - b_t)^2
\right]
=
\eta^\top
\mathbf{M}_t
\eta
-
2 b_t \mathbf{c}_t^\top \eta
+
b_t^2,
\end{equation}
where 
\begin{equation}\label{eq:2ndmoment}
\mathbf{M}_t
:=
\mathbb{E}_{\theta}
\!\left[
\mathbf{G}_t(\theta)^\top
a_t a_t^\top
\mathbf{G}_t(\theta)
\right],
\ \ 
\mathbf{c}_t
:=
\mathbb{E}_{\theta}
\!\left[
\mathbf{G}_t(\theta)^\top
a_t
\right].
\end{equation}
Using the above precomputations for the condensed matrices, the original SMPC problem \eqref{eq:smpc_problem} can be reformulated as 

\noindent\rule{\columnwidth}{0.4pt}
\textbf{Condensed PCE-SMPC}
{\small
\begin{subequations}\label{eq:det_smpc_problem}
\begin{align}
\min_{U} \quad 
& U^\top H U + 2 g^\top U
\label{eq:det-cost}
\\
\text{s.t.}\quad
& x^{\min} \leq \bar{\mathbf{E}}_t z_0 + \bar{\mathbf{F}}_t U \leq x_{t}^{\max} ,
\quad t=1,\dots,H,
\\
& \eta^\top \mathbf{M}_t \eta - 2 b_t \mathbf{c}_t^\top \eta + b_t^2
\le c_t^2,
\quad t=1,\dots,H,
\label{eq:det-2ndmoment}
\\
&  \mathbf1^H \otimes u_{\min}\leq U \leq \mathbf1^H \otimes u_{\max}
\end{align}
\end{subequations}
}

\vspace{-0.3cm}

\noindent\rule{\columnwidth}{0.4pt}

By exploiting the linear structure of the PPKO model and using the condensed matrices, the reformulated problem \eqref{eq:det_smpc_problem} has a dimension depending only on the control dimension and horizon, independent of the number of PCE polynomials and Koopman Observables in the PPKO. This approach also avoids SGPs, which typically introduce truncation errors and lead to large extended state-space models of PCE coefficients. The only approximation in this method arises from the quadrature used to evaluate the matrix-valued expectations in \eqref{eq:pcecost}–\eqref{eq:2ndmoment}, and the approximation error can be reduced offline to arbitrary accuracy by increasing the number of quadrature points.

\section{Numerical Examples}

This section provides numerical results demonstrating the performance of the proposed PPKO-SMPC framework on two nonlinear dynamical systems with parametric uncertainty: a Duffing oscillator and a reaction network in a continuous stirred-tank reactor (CSTR). These examples are selected because they exhibit strong nonlinear dynamics and pose nontrivial closed-loop regulation challenges. The PPKO-SMPC framework and numerical results can be reproduced with the code found in \url{https://github.com/Efstathios-Iliakis/Koopman-NN}.All computational examples were carried out on a Dell Latitude 7440 laptop equipped with a 13th Gen Intel Core i7-1365U processor and 32 GB of RAM, running Microsoft Windows 11 Enterprise.

\subsection{Stochastic Duffing Oscillator}
\label{subsec:duffing_setup}
Consider the controlled Duffing oscillator:
\begin{subequations}\label{eq:duffing_problem}
\begin{align}
\dot{x}_1 &= x_2 \label{eq:duffing_x1}\\
\dot{x_2} &= -\delta x_2 - x_1(\beta +\alpha x_1^2) + u \label{eq:duffing x_2}
\end{align}
\end{subequations}
where $x = [x_1\;x_2]^\top \in \mathbb{R}^2$ denotes the states and $u \in \mathbb{R}$ is the control input. The parameters $\delta \sim \mathcal{U} [0,1) $, $\beta \sim \mathcal{U} [-2,2)$, and $\alpha \sim \mathcal{U} [0,2)$ are uncertain parameters following uniform distributions.

\subsubsection{Simulations and training of PPKO}

The continuous-time dynamics  (\ref{eq:duffing_problem}) are discretized using the fourth order Runge-Kutta method with a sampling interval of $\Delta t = 0.02$ over 200 total time steps. To train the model, 20 distinct sets of parameter realizations are utilized. For each set, 20 different initial conditions are sampled from  uniform distributions $\mathcal{U}[-2, 2) \times \mathcal{U}[-2, 2)$ to complete the training dataset. The control inputs are sampled from a standard normal distribution and applied over the time horizon used during the training phase.

The Koopman observables are learned using the EDMD-DL method. The feature network is implemented as a two-layer fully connected neural network with 64 nodes per hidden layer and tanh activation functions, and is trained for up to 1000 epochs using mini-batches of size 2048. The polynomial matrices $A(\theta)$ and $B(\theta)$ for the PPKO model were approximated by multivariate Legendre polynomials up to the second degree. The PPKO model is learned via the alternating minimization procedure, where the coefficient matrices $A_i$ and $B_i$ were updated after each epoch with the closed form solution of the ordinary least squares problem. 
Early stopping with patience 100 was adopted. 

\begin{figure}
    \centering
    
    \vspace{0.2cm}
    \includegraphics[width=\columnwidth]{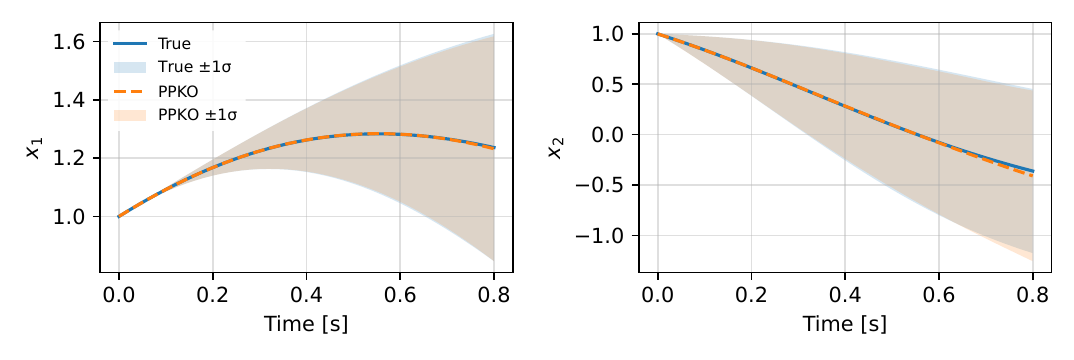} 

    \vspace{-0.3cm}
    
    \caption{Open-loop Duffing oscillator under parametric uncertainty for a  representative set of initial conditions $x_0 = [1,1]^\top$: mean trajectories with $\pm1\sigma$ bands for the true system and the PPKO model. Trajectories are generated from 30,000 random parameter samples $[\delta, \alpha, \beta]$, input profiles, and initial conditions over a 40-step horizon ($\Delta t = 0.02$).}
\label{fig:phase_portraits}
\end{figure}

 To assess the accuracy of the proposed PPKO model in open-loop prediction, the uncontrolled Duffing oscillator was propagated over a horizon of $H=40$ steps with $\Delta t = 0.02$ for several sets of initial conditions. The PPKO mean and variance were compared to Monte~Carlo (MC) ground truth ($N_{\mathrm{mc}} = 30{,}000$ samples) (e.g., Fig.~\ref{fig:phase_portraits}); the PPKO predictions closely match the MC reference in both the mean trajectory and the uncertainty envelope, confirming that the learned Koopman model accurately captures the stochastic open-loop dynamics.

\subsubsection{SMPC of stochastic Duffing oscillator}

\label{subsec:duff_control}

The PCE-SMPC controller \eqref{eq:det_smpc_problem} using the PPKO model was then implemented in the receding-horizon form with a prediction horizon of $H=5$. The Gauss--Legendre quadrature rule for the tensor-product with 5 nodes per uncertain parameter was used to evaluate the condensed matrices \eqref{eq:pcecost}–\eqref{eq:2ndmoment}. The stage-state weighting matrix was chosen as \(Q=\mathrm{diag}(5,2)\), the terminal weighting matrix was \(Q_f=\mathrm{diag}(200,120)\), and the input penalty was \(R=0.05I\). The SMPC dynamic optimization  \eqref{eq:det_smpc_problem} was solved at each step using CVXPY \cite{diamond2016cvxpy, agrawal2018rewriting} with the CLARABEL solver \cite{Clarabel_2024} and warm-started by shifting the previous solution forward by one step. The true system was propagated with a fourth-order Runge–Kutta scheme at a sampling time of $ \Delta t = 0.02$ s.

\begin{figure*}
    \centering
    
    \vspace{0.2cm}
    
    \includegraphics[width = 0.75\linewidth]{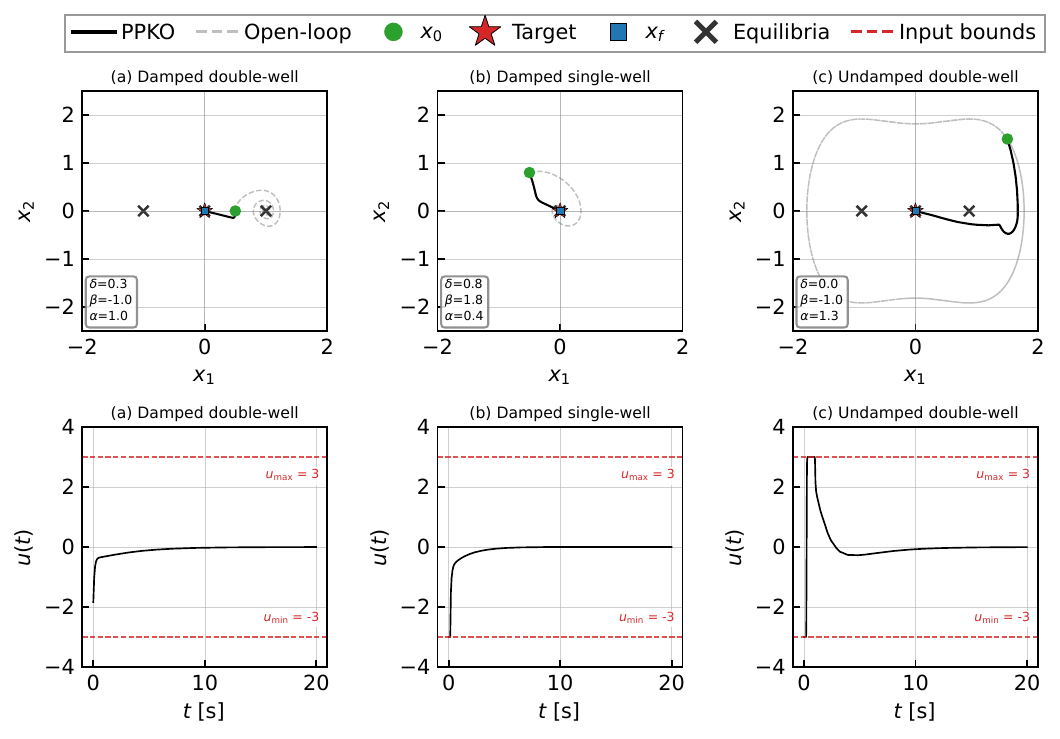}

    \vspace{-0.3cm}
    
    \caption{Closed-loop dynamics and control input profiles for 3 representative parametric uncertainty realizations corresponding to three different open-loop dynamical regimes: damped double-well (left), damped single-well (middle), and undamped double-well (right). Uncontrolled open-loop trajectories (dotted) are shown for comparison. $x_0$ is the initial point, $x_f$ is the final point of the simulation, and the target is the origin (0,0). Equilibria correspond to either a stable point or  saddle point.}
    \label{fig:control_duff}
\end{figure*}

To evaluate the robustness of our proposed framework to parametric uncertainty, three representative parameter configurations were selected, each producing qualitatively distinct dynamical behavior. 
In every case, the proposed stochastic Koopman–MPC scheme successfully steers the state to a neighborhood of the origin despite parametric uncertainty, demonstrating that the learned Koopman model captures sufficient dynamic structure to enable closed-loop stabilization across all three regimes. For qualitative comparison, uncontrolled open-loop trajectories and numerically computed equilibria along the line $x_2 = 0$ are overlaid in the phase portraits of Fig.\ \ref{fig:control_duff}.

\subsection{Control of Product Concentration in a Reaction Network}
\begin{figure}
\includegraphics[scale=0.75]{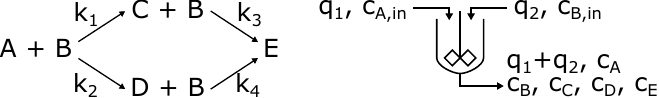}
\centering

\vspace{-.25cm}

\caption{Series-parallel reaction network and CSTR configuration studied (from \cite{von2020stochastic}).}
\label{fig: reaction scheme}
\end{figure}
We further evaluate the effectiveness of our PPKO models for the control of the concentration of a reaction network intermediate in a CSTR. The case study is adapted from a series–parallel nucleophilic aromatic substitution reaction network where the desired intermediate concentration is reduced by side-product formation in parallel reactions \cite{reizman2012automated}. Figure \ref{fig: reaction scheme} provides an illustration of the reaction network and the CSTR, and the governing equations for the CSTR are
\begin{small}
\begin{subequations}\label{eq: CSTR ODE}
\begin{align}
\frac{\mathrm{d}c_\mathrm{A}}{\mathrm{d}t} &= \frac{q_1}{V} c_\mathrm{A,in} - c_\mathrm{A} \frac{q_1 + q_2}{V} - c_\mathrm{A} c_\mathrm{B} (k_1 + k_2),  \\
\frac{\mathrm{d}c_\mathrm{B}}{\mathrm{d}t} &=   \frac{q_2}{V} c_\mathrm{B,in} -  c_\mathrm{B}  \frac{q_1 + q_2}{V} - c_\mathrm{A} c_\mathrm{B} (k_1 + k_2)  \\&\quad- c_\mathrm{B} c_\mathrm{C} k_3 -  c_\mathrm{B} c_\mathrm{D} k_4,  \\
\frac{\mathrm{d}c_\mathrm{C}}{\mathrm{d}t} &= -  c_\mathrm{C}  \frac{q_1 + q_2}{V} + c_\mathrm{A} c_\mathrm{B} k_1 - c_\mathrm{B} c_\mathrm{C} k_3, \\
\frac{\mathrm{d}c_\mathrm{D}}{\mathrm{d}t} &= -  c_\mathrm{D}  \frac{q_1 + q_2}{V} + c_\mathrm{A} c_\mathrm{B} k_2 - c_\mathrm{B} c_\mathrm{D} k_4,
\end{align}
\end{subequations}
\end{small}
where $c_i$ is the concentration of species $i=\{\mathrm{A},\mathrm{B},\mathrm{C},\mathrm{D}\}$, $c_{i,\mathrm{in}}$ is the inlet stream concentration of $c_i$, $q_1$ and $q_2$ are the total volumetric flow rates of inlet streams 1 and 2 which contain feed species $A$ and $B$ respectively, and $k_j$ is the rate constant for reaction $j=\{1,2,3,4\}$. The rate constants $k_1 \sim \mathcal{U}[0.2789, 0.8927]$ and $k_2 \sim \mathcal{U}[0.1894, 0.9331] $ are uniformly distributed uncertain parameters. The manipulated variable is $q_1$ which contains species $A$, and the controlled variable in the closed-loop simulations is $c_C$. Unlike the previous case study, the manipulated input enters the reactor dynamics in both a control-affine and nonlinear manner, which makes the control problem more challenging for linear Koopman predictors. All the parameters in the governing equations for this simulation are in Table 2 of \cite{tan2025offset}.

\subsubsection{Simulations and training of PPKO}
\begin{figure}
    \centering
    \includegraphics[width=1\linewidth]{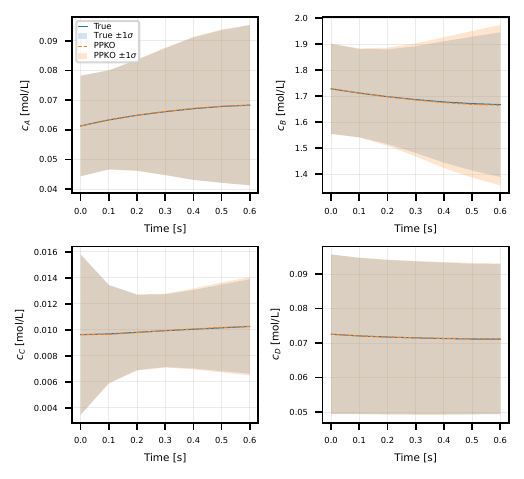}

    \vspace{-0.3cm}
    
   \caption{Open-loop CSTR dynamics under parametric uncertainty: mean trajectories with $\pm 1\sigma$ bands for the true system and the PPKO model. Trajectories are generated from random parameter samples $[k_1,k_2]$, input profiles, and initial conditions over a 10-step horizon ($\Delta t = 0.1$).}
    \label{fig: cstropenloop}
\end{figure}

The PPKO models were trained following a similar procedure with some minor modifications. First, the PCE coefficients of the matrices were obtained by solving a ridge regression problem with $\ell_2 = 10^{-5}$. Then, a key difference in this case study is that the steady-state equilibrium point varies with the uncertain parameters $k_1$ and $k_2$. To account for this variation, the model was trained using deviation variables relative to the corresponding parameter-dependent steady state. 
To evaluate the accuracy of our PPKO models in open-loop simulations, simulations were performed for different initial conditions, control trajectories, and uncertain parameter realizations over a horizon of $H=10$ time steps with $\Delta t=0.1$. Figure \ref{fig: cstropenloop} shows the trajectory envelopes across various parametric uncertainty realizations, illustrating that our PPKO model was able to accurately reproduce the open-loop CSTR dynamics.

\subsubsection{SMPC of CSTR}
Similar to the previous case study, the PCE-SMPC controller \eqref{eq:det_smpc_problem} using the PPKO model was  implemented in receding-horizon form with a prediction horizon of $H=10$. The tensor-product Gauss--Legendre quadrature rule with 4 nodes per uncertain parameter was used to evaluate the condensed matrices described in \eqref{eq:pcecost}–\eqref{eq:2ndmoment}. Since the control problem is to regulate $c_C$ only, the stage-state weighting matrix and terminal weighting matrix were chosen as \(Q=Q_f=\mathrm{diag}(0,0,1,0)\), and the input penalty was \(R=0.01I\). Similarly, the SMPC dynamic optimization  \eqref{eq:det_smpc_problem} was solved at each step using CVXPY \cite{diamond2016cvxpy, agrawal2018rewriting} with the OSQP solver \cite{osqp} and warm-started by shifting the previous solution forward by one step. The true system was propagated with a fourth-order Runge–Kutta scheme with a step size of $\Delta t = 0.01$ s.

\begin{figure}
    \centering
    \includegraphics[width=0.75\linewidth]{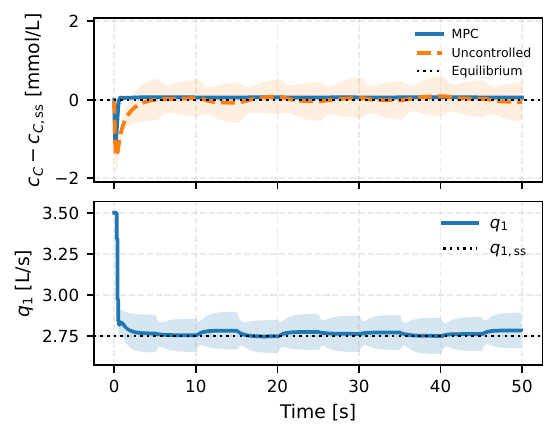}

    \vspace{-0.35cm}
    
   \caption{Closed-loop CSTR regulation under parametric uncertainty. Top: mean trajectories with $\pm 1\sigma$ bands for the PPKO model and uncontrolled model. Bottom: Mean profile and $\pm 1\sigma$ bands for the control input $q_1$. }
    \label{fig: cstrclosedloop}
\end{figure}

To evaluate the robustness of our proposed framework to parametric uncertainty, we performed a closed-loop simulation where disturbances were added to the feed concentration $c_{B,\text{in}}$ and the control task is to regulate $c_{C}$ back to the steady-state concentration $c_{C,\text{ss}}$. The simulation was performed with 100 randomly generated different parametric uncertainty realizations and control disturbance profiles. Figure \ref{fig: cstrclosedloop} shows the deviation trajectory and control input profile envelopes for the closed-loop simulations. The uncontrolled trajectory envelopes where the disturbances are not rejected by the controller are also displayed for direct comparison. In every case, the proposed stochastic Koopman-MPC scheme using the PPKO models is able to regulate $c_{C}$ to the steady state, demonstrating efficacy of our framework in closed-loop stabilization. 

\begin{figure}
    \centering
        \includegraphics[width=0.75\columnwidth]{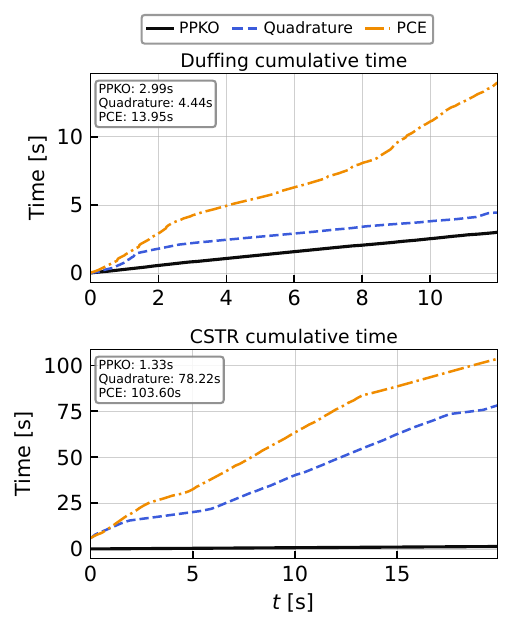} 
    \vspace{-0.25cm}
    
    \caption{Solve time comparison between 3 different SMPC schemes for Duffing oscillator (top) and the CSTR (bottom) case studies.}
\label{fig:solvetimeduffing}
\end{figure}

\subsection{Solve Time Comparison}
To evaluate the computational efficiency of the proposed framework, the cumulative solve times of PPKO are compared against Quadrature SMPC and PCE-SMPC in Fig.~\ref{fig:solvetimeduffing} across both benchmark systems for a single case. Quadrature SMPC is a scenario-based approximation of SMPC with the nonlinear model \eqref{eq:nonlinear} where expectations are evaluated using Gaussian quadrature, and PCE-SMPC is an intrusive PCE SMPC formulation where the stochastic Galerkin projections are evaluated by Gaussian quadrature. Full implementation details and source code are available at the repository linked at the beginning of this section. In both cases, without losing accuracy in the proposed control sequences, PPKO completes the closed-loop simulation faster than either baseline, with the advantage becoming substantially more pronounced on the higher-dimensional CSTR: here, PPKO achieves speed-up factors exceeding $50\times$ relative to both alternatives. This scaling behavior is a direct consequence of the condensing procedure, which reduces the online PPKO problem to a QP of dimension $Hn_u$, independent of both the lifted state dimension and the number of PCE basis terms. In contrast, the Quadrature and PCE-SMPC formulations require solving nonlinear programs whose size grows with the number of quadrature nodes or polynomial chaos coefficients, resulting in a computational cost that scales unfavorably with system dimension.

\section{Conclusion}
This article proposes a PCE-based Koopman operator framework for SMPC, where the Koopman operator is parametrized by PCEs. The model learns an approximately invariant subspace of observables from the EDMD-DL method, which provides a linear representation of the dynamics across various uncertain parameter realizations. This PCE-based structure enables the reformulation of the SMPC problem as a deterministic convex optimization after condensing. The dimension of these problems depends only on the control horizon and input dimension, and is independent of both the number of Koopman observables and PCE terms. Future work aims to explicitly quantify the approximation capabilities of the polynomial parametric Koopman operators (PPKO) \cite{guo2025learning}, integrate multi-step learning strategies \cite{wu2026least}, and incorporate robustness and closed-loop stability guarantees into our formulation \cite{zhang2022robust,strasser2024koopman}. 
\bibliography{ref}

@article{xiu2002wiener,
  title={The {Wiener}--{Askey} Polynomial Chaos for Stochastic Differential Equations},
  author={Xiu, D. and Karniadakis, G. E.},
  journal={SIAM Journal on Scientific Computing},
  volume={24},
  number={2},
  pages={619--644},
  year={2002}
}

@article{tan2025offset,
  title={Offset-free stochastic quadratic dynamic matrix control formulations using polynomial chaos expansions},
  author={Tan, Wallace Gian Yion and Ganko, Krystian and Santra, Srimanta and von Andrian, Matthias and Braatz, Richard D},
  journal={Control Engineering Practice},
  volume={165},
  artnum={106514},
  pages={106514},
  year={2025},
  publisher={Elsevier}
}

@article{zhang2022robust,
  title={Robust tube-based model predictive control with {K}oopman operators},
  author={Zhang, Xinglong and Pan, Wei and Scattolini, Riccardo and Yu, Shuyou and Xu, Xin},
  journal={Automatica},
  volume={137},
  pages={110114},
  year={2022},
  publisher={Elsevier}
}

@article{wu2026least,
  title={Least-Squares Multi-Step {Koopman} Operator Learning for Model Predictive Control},
  author={Wu, Liang and Tan, Wallace Gian Yion and Zhou, Leqi and Braatz, Richard D and Drgona, Jan},
  journal={arXiv preprint arXiv:2601.11901},
  year={2026}
}

@article{mishra2024polynomial,
  title={Polynomial chaos-based stochastic model predictive control: {An} overview and future research directions},
  author={Mishra, Prabhat K. and Paulson, Joel A. and Braatz, Richard D.},
  journal={arXiv preprint arXiv:2406.10734},
  year={2024}
}

@article{strasser2024koopman,
  title={{Koopman}-based feedback design with stability guarantees},
  author={Str{\"a}sser, Robin and Schaller, Manuel and Worthmann, Karl and Berberich, Julian and Allg{\"o}wer, Frank},
  journal={IEEE Transactions on Automatic Control},
  volume={70},
  number={1},
  pages={355--370},
  year={2024},
  publisher={IEEE}
}

@article{reizman2012automated,
    title={An Automated Continuous-Flow Platform for the Estimation of Multistep Reaction Kinetics},
    author={Reizman, Brandon J. and Jensen, Klavs F.},
    journal={Organic Process Research \& Development},
    volume={16},
    number={11},
    pages={1770--1782},
    year={2012},
    publishers={ACS Publications},
    notes = {{Publisher}: American Chemical Society Publications}
}

@article{paulson2020stochastic,
  title={Stochastic model predictive control with joint chance constraints},
  author={Paulson, Joel A. and Buehler, Edward A. and Braatz, Richard D. and Mesbah, Ali},
  journal={International Journal of Control},
  volume={93},
  number={1},
  pages={126--139},
  year={2020},
  publisher={Taylor \& Francis}
}

@inproceedings{von2020stochastic,
    title={Stochastic Dynamic Optimization and Model Predictive Control Based on Polynomial Chaos Theory and Symbolic Arithmetic},
    author={von Andrian, Matthias and Braatz, Richard D},
    booktitle={Proceedings of the American Control Conference},
    pages={3399--3404},
    year={2020},
    organizations={IEEE}
}

@article{williams2016extending,
  title={Extending data-driven {K}oopman analysis to actuated systems},
  author={Williams, Matthew O and Hemati, Maziar S and Dawson, Scott TM and Kevrekidis, Ioannis G and Rowley, Clarence W},
  journal={IFAC-PapersOnLine},
  volume={49},
  number={18},
  pages={704--709},
  year={2016},
  publisher={Elsevier}
}

@article{dai2016distributed,
  title={Distributed stochastic {MPC} of linear systems with additive uncertainty and coupled probabilistic constraints},
  author={Dai, Li and Xia, Yuanqing and Gao, Yulong and Cannon, Mark},
  journal={IEEE Transactions on Automatic Control},
  volume={62},
  number={7},
  pages={3474--3481},
  year={2016},
  publisher={IEEE}
}

@article{bernardini2011stabilizing,
  title={Stabilizing model predictive control of stochastic constrained linear systems},
  author={Bernardini, Daniele and Bemporad, Alberto},
  journal={IEEE Transactions on Automatic Control},
  volume={57},
  number={6},
  pages={1468--1480},
  year={2011},
  publisher={IEEE}
}

@article{ruppen1995optimization,
  title={Optimization of batch reactor operation under parametric uncertainty—{Computational} aspects},
  author={Ruppen, D. and Benthack, C. and Bonvin, D.},
  journal={Journal of Process Control},
  volume={5},
  number={4},
  pages={235--240},
  year={1995},
  publisher={Elsevier}
}

@article{mammarella2018offline,
    title={An Offline-Sampling {SMPC} Framework With Application to Autonomous Space Maneuvers},
    author={Mammarella, Martina and Lorenzen, Matthias and Capello, Elisa and Park, Hyeongjun and Dabbene, Fabrizio and Guglieri, Giorgio and Romano, Marcello and Allg{\"o}wer, Frank},
    journal={IEEE Transactions on Control Systems Technology},
    volume={28},
    number={2},
    pages={388--402},
    year={2018},
    publishers={IEEE},
    notes = {{Publisher}: Institute of Electrical and Electronics Engineers}
}

@article{li2017extended,
  title={Extended dynamic mode decomposition with dictionary learning: {A} data-driven adaptive spectral decomposition of the {K}oopman operator},
  author={Li, Qianxiao and Dietrich, Felix and Bollt, Erik M and Kevrekidis, Ioannis G},
  journal={Chaos: An Interdisciplinary Journal of Nonlinear Science},
  volume={27},
  number={10},
  pages = {103111},
  year={2017},
  publisher={AIP Publishing}
}

@article{korda2018linear,
  title={Linear predictors for nonlinear dynamical systems: {K}oopman operator meets model predictive control},
  author={Korda, Milan and Mezi{\'c}, Igor},
  journal={Automatica},
  volume={93},
  pages={149--160},
  year={2018},
  publisher={Elsevier}
}

@article{proctor2018generalizing,
  title={Generalizing {K}oopman theory to allow for inputs and control},
  author={Proctor, Joshua L and Brunton, Steven L and Kutz, J Nathan},
  journal={SIAM Journal on Applied Dynamical Systems},
  volume={17},
  number={1},
  pages={909--930},
  year={2018},
  publisher={SIAM}
}

@article{guo2025learning,
  title={Learning parametric {K}oopman decompositions for prediction and control},
  author={Guo, Yue and Korda, Milan and Kevrekidis, Ioannis G and Li, Qianxiao},
  journal={SIAM Journal on Applied Dynamical Systems},
  volume={24},
  number={1},
  pages={744--781},
  year={2025},
  publisher={SIAM}
}

@article{mesbah2016stochastic,
  title={Stochastic model predictive control: {A}n overview and perspectives for future research},
  author={Mesbah, Ali},
  journal={IEEE Control Systems Magazine},
  volume={36},
  number={6},
  pages={30--44},
  year={2016},
  publisher={IEEE}
}

@ARTICLE{10683964,
  author={Wu, Liang and Braatz, Richard D.},
  journal={IEEE Transactions on Automatic Control}, 
  title={A {D}irect {O}ptimization {A}lgorithm for {I}nput-{C}onstrained {MPC}}, 
  year={2025},
  volume={70},
  number={2},
  pages={1366-1373}
  }

@ARTICLE{11240592,
  author={Wu, Liang and Xiao, Wei and Braatz, Richard D.},
  journal={IEEE Transactions on Automatic Control}, 
  title={E{IQP}: {E}xecution-time-certified and Infeasibility-detecting {QP} Solver}, 
  year={2025},
  volume={},
  number={},
  pages={1-16},
  doi={10.1109/TAC.2025.3631575}
  }

@ARTICLE{11431115,
  author={Wu, Liang and Braatz, Richard D.},
  journal={IEEE Transactions on Automatic Control}, 
  title={A {Q}uadratic {P}rogramming {A}lgorithm with ${O}(n^3)$ {T}ime {C}omplexity}, 
  year={2026},
  volume={},
  number={},
  pages={1-16},
  doi={10.1109/TAC.2026.3673192}
  }

@article{kim2025ksmpc,
  title={{K-SMPC}: {K}oopman operator-based stochastic model predictive control for enhanced lateral control of autonomous vehicles},
  author={Kim, Jin Sung and Quan, Ying Shuai and Chung, Chung Choo and Choi, Woo Young},
  journal={IEEE Access},
  year={2025},
  pages = {13944--13958},
  volume = 13,
  publishers={IEEE},
  notes={Early access}
}

@article{Clarabel_2024,
  title={Clarabel: An interior-point solver for conic programs with quadratic objectives},
  author={Paul J. Goulart and Yuwen Chen},
  year={2024},
    journal={arXiv preprint arXiv:2405.12762},
  eprint={2405.12762},
  archivePrefix={arXiv},
  primaryClass={math.OC}
}

@article{diamond2016cvxpy,
  author  = {Steven Diamond and Stephen Boyd},
  title   = {{CVXPY}: {A} {P}ython-embedded modeling language for convex optimization},
  journal = {Journal of Machine Learning Research},
  year    = {2016},
  volume  = {17},
  number  = {83},
  pages   = {1--5},
}

@article{agrawal2018rewriting,
  author  = {Agrawal, Akshay and Verschueren, Robin and Diamond, Steven and Boyd, Stephen},
  title   = {A rewriting system for convex optimization problems},
  journal = {Journal of Control and Decision},
  year    = {2018},
  volume  = {5},
  number  = {1},
  pages   = {42--60},
}

@article{osqp,
  author  = {Stellato, B. and Banjac, G. and Goulart, P. and Bemporad, A. and Boyd, S.},
  title   = {{OSQP}: {An} operator splitting solver for quadratic programs},
  journal = {Mathematical Programming Computation},
  volume  = {12},
  number  = {4},
  pages   = {637--672},
  year    = {2020},
  dois     = {10.1007/s12532-020-00179-2},
  urls     = {https://doi.org/10.1007/s12532-020-00179-2},
}
\bibliographystyle{IEEEtran} 
\end{document}